\documentclass[12pt]{article}

\usepackage{graphicx,amsmath,amssymb,color,bm}

\begin{document}

\centerline {\Large\textbf {Anomalous magneto-transport properties of bilayer phosphorene
}}

\centerline {\Large \textbf {}}\vskip0.6 truecm

\centerline{Jhao-Ying Wu$^{1,*}$, Wu-Pei Su$^{2}$, Godfrey Gumbs$^{3}$}

\centerline{$^{1}$Center of General Studies, National Kaohsiung University of Science and Tachnology, Kaohsiung, Taiwan 811}
\centerline{$^{2}$Department of Physics, University of Houston, Houston, Texas}
\centerline{$^{3}$Department of Physics and Astronomy, Hunter College at the City University of New York,}
\centerline{New York, New York 10065, USA}

\begin{abstract}

The magneto-transport properties of phosphorene are investigated by employing the  generalized tight-binding model to calculate the energy bands. For bilayer phosphorene, a composite magnetic and electric field is shown to induce a feature-rich Landau level (LL) spectrum which includes two subgroups of low-lying LLs. The two subgroups possess distinct features in level spacings, quantum numbers, as well as field dependencies. These together lead to anomalous quantum Hall (QH) conductivities which include a well-shape, staircase and composite quantum structures with steps having varying heights and widths. The Fermi energy-magnetic field-Hall conductivity ($E_{F}-B_{z}-\sigma_{xy}$)  and Fermi energy-electric field-Hall conductivity ($E_{F}-E_{z}-\sigma_{xy}$) phase diagrams clearly exhibit  oscillatory behaviors and cross-over from integer to half-integer QH effect. The predicted results should be verifiable by magneto-transport measurements in a dual-gated system.

\vskip0.6 truecm

\noindent
%PACS:\ \  {\bf 71.70.Di,31.15.ar,71.30.h}

\end{abstract}

\newpage

\bigskip

\centerline {\textbf {I. INTRODUCTION}}

Two-dimensional (2D) layered systems, having nano-scaled thickness and unique geometric symmetries, have found themselves in the main stream of material sciences especially after the success of graphene. These novel materials have been found to possess critical applications due to their transport and optical properties. Other 2D layered structures include silicene \cite{Takagi:2015}, germanene \cite{Derivaz:2015}, boron nitride (BN) \cite{ZK2017}, gallium nitride (GaN) \cite{ALB2016}, transition-metal dichalcogenides (TMDs) \cite{Wang2012}, and stanene \cite{Zhu2015,YJ2018}.  However, they have certain shortcomings, like zero/narrow band-gaps in IV-group materials \cite{Balendhran2015}, and low carrier mobility for some TMDs. This deficiency can greatly limit their application and deter progress in electronic devices. On the other hand, V-group elements have been attracting a considerable amount of  attention due to their moderate band gap and carrier mobility.Among these elements, few-layer black phosphorus (phosphorene) has been fabricated successfully by using mechanical cleavage \cite{PLiu2014,Taimur2017}, liquid exfoliation \cite{Pkang2015,Backes2017}, and mineralizer-assisted short-way transport reaction \cite{PNilges2008,PKopf2014}.

Few-layer phosphorene inherently has an energy band gap of $\sim 0.5-2$ eV \cite{gap1,PRudenko,gap}, as deduced from optical measurements \cite{PLiu2014,Zhang2014}. Such a gap is larger than that ($\sim0.3$ eV) for its bulk counterpart  \cite{gap1,Li2014,Han2014}. Transport measurements show that phosphorene-based field-effect transistor exhibits an on/off ratio of $10^{5}$ and a carrier mobility at room temperature as high as $10^{3}$ cm$^{2}/$V.s \cite{Li2014,Ma2017,Yang2018}. Additionally, layered black phosphorus systems display unusual energy spectra and quantum Hall effect (QHE) due to magnetic quantization \cite{Likai2016,Lado2016,Jiang2016,Gen2017,Fangyuan2018}. Few-layer black phosphorus is expected to have unparalleled potential in the next-generation electronic devices \cite{Tiana2016,HX2018,HZ2018}.

Each phosphorene layer possesses a puckered structure, mainly due to the $sp^{3}$ hybridization of ($3s,3p_{x},3p_{y},3p_{z}$) orbitals. This deformed hexagonal lattice in the x-y plane is quite different from the honeycomb lattice of the group-IV systems \cite{Rudenko2014}. That unique geometric structure fully dominates the low-lying energy bands which are highly anisotropic, e.g., the linear and parabolic dispersions near the Fermi energy $E_{F}$, respectively, along  the $\widehat{k_{x}}$ and $\widehat{k_{y}}$ directions \cite{PRudenko}. The anisotropic behaviors are clearly revealed in other physical properties, as verified by recent measurements of the optical and transport properties \cite{Li2014,TLow,Qiao2014}. This provides a unique advantage for phosphorene in comparison with Mo$S_2$- and related semiconductors. The unusual anisotropy could be utilized in the design of unconventional thermoelectric devices. For example, a thermal gradient and a potential difference could be applied in two orthogonal directions, leading to one having higher thermal conductivity and another with larger electrical conductivity \cite{Fei2014,LingX2015}. Moreover, this intrinsic property will greatly diversify the quantization phenomena.

Over the years, there have been several experimental and theoretical studies of the rich and unique quantum transport properties of graphene and -related systems. Magnetic transport measurements on monolayer graphene have revealed the unconventional half integer Hall conductivity $\sigma_{xy}=4(m + 1/2)e^{2}/h$, where $m$ is an integer and the factor of $4$ is present due to  spin and sublattice degeneracy. This unusual quantization is attributed to the quantum anomaly of the $n=0$ Landau level (LL) corresponding to the Dirac point.  Regarding AB-stacked bilayer graphene, the Hall conductivity is confirmed to be $\sigma_{xy}$=$4m^{'}e^{2}/h$ ($m^{\prime}$ a non-zero integer). Furthermore, there exists an unusual integer quantum Hall conductivity, a double step of $\sigma_{xy}$ = 8$e^{2}$/h, at zero energy and low magnetic field \cite{Novoselov2006}. This mainly comes from the $n=0$ and $n=1$ LLs in the first group \cite{Lai2008}. The foregoing  interesting phenomena of electronic transport properties of graphene opens the door to exploring the configuration-enriched QHE in other novel 2D materials.

In this paper, the Kubo formula based on linear response theory, in conjunction with the generalized tight-binding model, is employed to investigate the unusual QHE of bilayer phosphorene under a uniform perpendicular electric field $E_{z}$. The relations between the geometrical structures, intrinsic interactions, electric field $E_{z}$ and magnetic field $B_{z}$ are investigated in detail. The model is very useful for identifying  the magneto-electronic selection rules in the static limit.  That is, the available transition channels in the magneto-transport properties could be examined thoroughly. The dependencies of quantum conductivity on the Fermi energy $E_{F}$, $E_{z}$ and $B_{z}$ are examined in detail. The close relations between the electronic structure, the LLs and the transition channels are also the object of our investigation.

By varying the electric field strength, we can greatly diversify the magnetic quantization and enrich the transport properties. The diverse features could cover half-integer and integer conductivities with distinct steps, a vanishing or non-zero conductivity at the neutral point, and the well-like, staircase, composite, and anomalous plateau structures in the field dependencies. The strong field-dependent characteristics mainly originate from  two-groups of LLs, which present many anticrossing/crossing behavior. This study shows that the feature-rich LLs can create extraordinary magneto-transport properties.

\bigskip
\bigskip
\centerline {\textbf {II. METHODS}}
\bigskip
\bigskip

Monolayer phosphorene, with a puckered honeycomb structure, has a primitive unit cell containing four phosphorous atoms, as depicted by the dashed yellow lines in Fig. 1(a). Two of the four phosphorous  atoms are located on the lower (the red circles) or higher (the green circles) sublattice sites.  Similar structures are revealed in few-layer systems, e.g, bilayer phosphorene as shown  in Fig. 1(b) for AB stacking. The low-lying energy bands are dominated by the interactions between the 3$p_{z}$ orbitals \cite{PRudenko}. The few-layer Hamiltonian is represented by

\begin{equation}
H=\sum_{i=1,l}^{4}(\varepsilon_{i}^{l}+U_{i}^{l})c_{i}^{l}c_{i}^{\dag\,l}+
\sum_{\langle i,j \rangle, l}h^{ll}_{ij}c_{i}^{l}c_{j}^{\dag\,l}+
\sum_{\langle i,j \rangle, l \neq l^{\prime}}h^{\prime\,ll^\prime}_{ij}c_{i}^{l}c_{j}^{\dag\,l^{\prime}} \  .
\end{equation}
In this notation, $\varepsilon_{i}^{l}$ is zero in a monolayer, but for a few-layer system, it is a layer- and sublattice-dependent site energy due to the chemical environment. Also, $U_{i}^{l}$ is the Coulomb potential energy induced by an electric field. They both contribute to the diagonal matrix elements. In the absence of a magnetic field, the summation is written as $\Sigma_{i=1,l}^{4}$, but in the presence of a magnetic field it becomes $\Sigma_{i=1,l}^{4R_{B}}$ (seen later).  Also, $c_{i}^{l}$ ($c^{\dag\,l^{\prime}}_{j}$) is an annihilation (creation) operator, $h^{ll}_{ij}$ and $h^{\prime\,ll^\prime}_{ij}$ are, respectively, the intralayer and interlayer hopping integrals, and the effective interactions used in the calculations cover the fourth and fifth neighboring atoms. These hopping parameters have been adopted from Ref. \cite{PRudenko}.

Black phosphorene is assumed to be in the presence of a uniform perpendicular magnetic field. The magnetic flux through a unit cell is $\Phi= a_{1}a_{2}B_z$, where $a_{1}=3.27$ ${\AA}$ and $a_{2}=4.43$ ${\AA}$ are lattice constants (the lattice vectors are shown as the yellow arrows in Fig. 1(a)). The vector potential, $\vec{A}=(B_z x)\hat{y}$, creates an extra magnetic Peierls phase of $\exp\{i[\frac{2\pi}{\phi_{0}}\int \vec{A}\cdot d\vec{r}] \}$, leading to a new period along $\hat x$ and thus an enlarged rectangular unit cell with ${4R_B= 4\phi_0\,/\Phi}$ atoms in one layer, as illustrated in Fig. 1(c). Here, $\phi_{0}$ ($=h/e=4.1\times 10^{-15}$ T$\cdot m^{2}$) is the magnetic flux quantum; ${\phi_0\,/\Phi}$ is chosen to be an integer. The reduced first Brilloun zone has an area of ${4\pi^2\,/a_1a_2R_B}$. For bilayer black phosphorous, the magnetic Hamiltonian matrix is very large with 8$R_{B}$$\times$8$R_{B}$ matrix elements within achievable experimental field strengths, e.g., the dimension of 16800 at ${B_z=30}$ T.

The spatial distributions of subenvelope functions derived from the generalized tight-binding model are utilized to characterize the magnetic quantum numbers and the types of LLs \cite{Wu2017, CSC2018, SPH2019}. They are useful for explaining the peculiar LL behaviors. To achieve the experimentally attainable field strengths, a band-like matrix has been developed to solve the huge matrix efficiently \cite{CYLin2015}. Accordingly, we can observe the strong electrically-tunable LL spectra when $B_{z}<60$ T. The method incorporates the intra-layer and inter-layer atomic interactions, and the effects due to external fields, simultaneously. It is applicable to the study of the quantization effect in arbitrarily stacked layered materials under any form of external fields. Moreover, the results are accurate and reliable within a wide energy range.

The magnetically quantized LLs can create unique transport properties. Within linear response theory, the transverse Hall conductivity is evaluated from the Kubo formula \cite{Dutta2012}:

\begin{equation}
\sigma_{xy}=\frac{ie^{2}\hbar}{S}\sum_{\alpha\neq\beta}(f_{\alpha}-f_{\beta})\frac{\langle\alpha|\mathbf{\dot{u_{x}}}|\beta\rangle\langle\beta|\mathbf{\dot{u_{y}}}|\alpha\rangle}{(E_{\alpha}-E_{\beta})^{2}} \ .
\end{equation}
Here, $|\alpha\rangle$ is a LL state with energy $E_{\alpha}$, S the area of the enlarged unit cell, $f_{\alpha,\beta}$ the Fermi-Dirac distribution functions, and $\mathbf{\dot{u_{x}}}$ ($\mathbf{\dot{u_{y}}}$) the velocity operator along $\hat{x}$ ($\hat{y}$). The matrix elements of the velocity operators, which determine the permitted inter-LL transitions, are evaluated by the gradient approximation \cite{Johnson1973}:

\begin{equation}
\begin{array}{cccc}
\langle\alpha|\mathbf{\dot{u_{x}}}|\beta\rangle=\frac{1}{\hbar}\langle\alpha|\frac{\partial H}{\partial k_{x}}|\beta\rangle, \\
\langle\alpha|\mathbf{\dot{u_{y}}}|\beta\rangle=\frac{1}{\hbar}\langle\alpha|\frac{\partial H}{\partial k_{y}}|\beta\rangle.
\end{array}
\end{equation}

The calculated procedure for a two-dimensional system is quite a contrast to that of a one-dimensional ribbon \cite{Lado2016,Jiang2016,Yuan2016}. In ribbons, the periodicity along the longitudinal direction (parallel to the edges) can be independent of the magnetic field in the Landau gauge. The quantum confinement would hinder the magnetic quantization, leading to a partially dispersionless energy spectra \cite{Chung2016} rather than the dispersionless LLs in a two-dimensional system \cite{Wu2017,CSC2018,SPH2019}. Our method warrants more reliable results on QHE for an arbitrary Fermi level and external field strength.

\bigskip
\bigskip
\centerline {\textbf {III. RESULTS AND DISCUSSION}}
\bigskip
\bigskip

The special lattice structure and complex hopping integrals generate rich band structures. Bilayer black phosphorous (BP) has a direct gap of ${E_g\approx\,1}$ eV near the $\Gamma$ point, as illustrated in  Fig. 1(d) by the black curves. The highly anisotropic energy bands yield the approximately linear and parabolic dispersions along $\Gamma$X and $\Gamma$Y (${\hat k_x}$ and ${\hat k_y}$), respectively. An electric field $E_{z}$ could reduce the energy gap considerably. The semiconductor-semimetal transition appears at a strength larger than ${E_{z,c}\simeq 0.3}$ ({V/{\AA}}; blue  curves), for which the valence and conduction bands are transformed into linearly intersecting bands and oscillatory bands along $\Gamma$Y and $\Gamma$X, respectively, highlighted in the upper inset. The extreme points remain at the $\Gamma$ point. Two Dirac points (saddle points) are situated on both sides of the $\Gamma$ point along ${+\hat k_y}$ and ${-\hat k_y}$ (${+\hat k_x}$ and ${-\hat k_x}$). Especially, two constant-energy loops, inner and outer, surround the $\Gamma$ point in the region between the extreme and saddle point energies. This is illustrated by the contour plot for $E^{c}$=0.04 eV in the lower inset. The electronic states near the Dirac and $\Gamma$ points are magnetically quantized into two distinct LL subgroups. The hybridization of the two subgroups reaches a maximum at the saddle point energy ($E^{c}\approx0.028$ eV; $E^{v}\approx-0.035$ eV), which could cause peculiar optical \cite{WJY2018} and transport properties.

All the critical points and constant-energy loops in the energy-wave vector space will dominate the main features of the LLs.  For $E_{z}<E_{z,c}$ (Figs. 2(a) and 2(b)), the states around the $\Gamma$ point contribute to the low-lying LLs, which are four-fold degenerate for each (${k_x,k_y}$) state including spin and localization-center degeneracies. The occupied LLs are not symmetrically located  with respect to the unoccupied ones near the Fermi level. The quantum number $n^{c}$ ($n^{v}$) for each conduction (valence) LL is clearly identified from the number of zero nodes in the wavefunctions. For example, the four low-lying conduction/valence LLs have $n^{c,v}$=0, 1, 2; 3. Though they have well-behaved $B_z$-dependence, their energies cannot be well described by a simple relation especially for higher energy and field strength. This is different from the square-root dependence in monolayer graphene \cite{JHHo2008}, and the linear dependence in AB-stacked bilayer graphene \cite{Lai2008} and $MoS_{2}$ \cite{HoY2014}, due to the highly anisotropic energy dispersion.

A modest value for $E_{z}$ would lower (raise) the lowest unoccupied LL (highest occupied LL) energy, as shown in Fig. 2(b). The drastic changes in the LL spectrum come about when $E_{z}$ is increased to over $E_{z,c}$ (Fig. 2(c)). First, the lowest unoccupied and the highest occupied LLs merge together due to the zero band gap. Then, the low-lying LLs evolve into two groups around $E_{F}$. The lower one, namely the D-group with double degeneracy, could be fitted by a square-root relation $\sqrt{n_D B_z}$. That corresponds to the magnetic quantization of electronic states near the two Dirac points, similar to graphene \cite{JHHo2008}. For the higher one, the $\Gamma$-group, beginning at the band-edge-state energies at $\Gamma$ point (indicated by the blue arrows), the level energies are inversely proportional to $B_z$. $\Gamma$-group LLs with their own set of quantum numbers $n_\Gamma$ are associated with the inner constant-energy loops (see the lower inset of Fig. 1(d)) and frequently cross the LLs from the outer loops. $\Gamma$- and D-group LLs have strong hybridizations around the saddle point energies, indicated by the green arrows, where they cross and anticross each other alternatively. The anticrossings clearly illustrate that such LLs are composed of multi-oscillation modes \cite{Wu2017,LCY2014, HYH2013}.

The strong electrically-tunable LL spectrum is demonstrated in Fig. 2(d) for $B_{z}$=30 T. The zero-energy LL is generated at $E_{z}>E_{z,c}\approx$ 0.305. The larger $n^{c,v}$ requires the stronger $E_{z}$ to form the D-group LL. That is, two entangled LLs gradually merge together with increasing $E_{z}$. Above the region, one sees that $\Gamma$-group LLs are formed beyond the turning points (the first turning point is indicated by the green arrow). The energies of $\Gamma$-group LLs increase with $E_{z}$ quickly. This distinguishes them from the other LLs coming from the outer constant-energy loops. Clearly, the feature-rich LL spectrum strongly depends on the energy regime as well as strengths of the external field. This could result in interesting magneto-transport properties.

The dependence of the Hall conductivity on the Fermi energy is very sensitive to $E_{z}$, as exhibited in Figs. 3(a) through 3(d) for $B_{z}$=30 T. When $E_{z}=0$ (Fig. 3(a)), the conductivity is quantized as $\sigma_{xy}=2me^{2}/h$ (where $m$ is an integer) with almost equal plateau widths. This reflects the parabolic energy dispersion at $B_{z}=0$. A wide plateau of $\sigma_{xy}=0$, representing the insulating characteristic, centers around zero energy. The strongly anisotropic selection rules of available interband inter-LL transitions \cite{WJY2018} are responsible for the intrinsic zero conductivity. This is different from the case in graphene, where the available interband inter-LL transition channels $\Delta n=1$ and $\Delta n=-1$, respectively, possess opposite contributions to $\sigma_{xy}$ and cancel each other \cite{DoTN2017}. The intraband inter-LL transitions determine the magnitude of $\sigma_{xy}$. That is, when $E_{F}$ is between the $n^{c}=n-1$ and $n^{c}=n$ LLs ($n$ is a positive integer), the transition between the two LLs mainly contributes to $\sigma_{xy}=+n$ (in units of $2e^{2}/h$), while $E_{F}$ between $n^{v}=n-1$ and $n^{v}=n$ LLs it leads to $\sigma_{xy}=-n$. Thus, varying $E_{F}$/passing the LLs (shown in the upper inset) makes the step structures. The width of the zero-conductivity plateau can be reduced by $E_{z}$ considerably, while for a modest value of $E_{z}$ the quantization relation $\sigma_{xy}=2me^{2}/h$ remains (Fig. 3(b)). It is until $E_{z}>E_{z,c}$ that $\sigma_{xy}$ displays distinct features, like in Fig. 3(c) for $E_{z}=0.32$. There are three main characteristics. First of all, the plateau of $\sigma_{xy}=0$ disappears, which is attributed to the creation of zero-energy LL. Secondly, around $E_{F}=0$ the quantized values of the Hall conductivity are changed from integer to half-integer with the double level degeneracy, i.e., from $\sigma_{xy}=2me^{2}/h$ to $\sigma_{xy}=4(m^{\prime}+1/2)e^{2}/h$ ($m^{\prime}$ a nonzero integer). This manifests the formation of D-group LLs. Thirdly, there exist a few of very narrow plateaus, as indicated by the green arrows. They arise from the lift of LL degeneracy or the appearances of $\Gamma$-group LLs (the purple line in the upper inset). Fig. 3(d) displays the region of further lower $E_{F}$. It demonstrates that the coexistence of two groups of LLs could induce the nonuniform plateau widths. However, when $E_{F}$ becomes smaller than the extreme-point energy at $\Gamma$ point ($\approx -0.053$ eV), the structures turn monotonous, i.e, they obey the relation $\sigma_{xy}=2me^{2}/h$. The regular quantization means that the LLs originating from the outer constant-energy loops (see the lower inset of Fig. 1(d)) become dominant.

The $B_{z}$-dependent Hall conductivity exhibits a variety of features when $E_{z}$ is varied, as shown in Figs. 4(a)-4(d). These characteristics are due to the behavior of the LL energy spectrum as a function of magnetic field seen in the insets of each panels. For $E_{z}<E_{z,c}$ (Fig. 4(a)), strengthening $B_{z}$ decreases the quantum number of the highest occupied LL and increases the inter-LL spacing of the dominant transition channel around $E_{F}$ (the dashed green line). Consequently, the height/width of the Hall plateau is reduced/broadened when passing through a LL. This is in marked contrast to the case shown in Figs. 3(a) and 3(b) where varying $E_{F}$ barely changes the plateau width. The value of $\sigma_{xy}$ drops to zero over a critical strength $B_{z,c}$ (not shown). The larger the value of $E_{F}$ is, the stronger strength of $B_{z,c}$ is required. As for $E_{z}>E_{z,c}$, we observe that there are three main types of $\sigma_{xy}$-$B_{z}$ dependencies for chosen $E_{F}$, as displayed in Figs. 4(b) through 4(d). Firstly, when $E_{F}$ is located in the linear energy region (the inset in Fig. 4(b)), the conductivity exhibits the half-integer QHE in units of $4e^{2}/h$. There is a narrow plateau at $B_{z}\approx 36$ T due to a pair of entangled $n^{c}_{D}=1$ LLs. $\sigma_{xy}$ does not achieve the zero value because of the existence of a zero energy LL. Secondly, $E_{F}$ between in the energies of extreme and saddle points as shown in the inset of Fig. 4(c), where two groups of LLs coexist, the conductivity possesses a composite quantum structure. The reason of this is that the intra inter-LL transitions of D- and $\Gamma$-group LLs contribute to positive and negative $\sigma_{xy}$, respectively. Therefore, with decreased quantum numbers of D-/$\Gamma$-group LLs, the value of $\sigma_{xy}$ is reduced/increased. This results in the unusual structures for certain strengths of $B_{z}$, including the double-width plateau (the red rectangle), bulging (the yellow rectangle) and dip (the green rectangle) structures. The first kind of structure happens when $E_{F}$ is located at a LL crossing point exactly, while the latter two occur when $E_{F}$ is slightly above and below a LL crossing point, respectively. Lastly, when $E_{F}$ is higher than the band-edge-state energy at the $\Gamma$ point (Fig. 4(d)), the monotonic staircase structure of integer QHE is similar to that of a 2D electron gas.

The dip and bulging structures appear frequently in $\sigma_{xy}$-$E_{z}$ plots, as shown in Figs. 5(a) and 5(b) for different $E_{F}$'s. The unusual staircase behavior begins when the $\Gamma$-group of LLs (in purple color) is formed at $E_{z}>E_{z,c}$, as seen in the insets. The higher value of $E_{F}$ is, the larger $E_{z,c}$ is needed. Contrary to the results determined by varying $B_{z}$, the quantum number of the highest occupied LL in each group increases with $E_{z}$. Therefore, the plateau height is increased/reduced passing a LL from the outer/inner constant-energy loops, leading to the well-like staircases. The widths of these particular structures are strongly governed by the LL-spacings/$B_{z}$-strength.

The $E_{F}$ versus $B_{z}$ phase diagram for the conductivity (color scale in units of $2e^{2}/h$) is helpful to give the whole picture of the feature-rich transport properties. With $E_{z}>E_{z,c}$ (Fig. 6(a)), the Hall plateaus appear as colored bands diverging from the charge neutrality point at $B_{z}\rightarrow 0$. For $B_{z}<10$ T, the quantized number $m$ around $E_{F}$=0 come in a regular order of odd ones, i.e., $m=1, 3, 5$, etc., in correspondence to the D-group of LLs. With the increase of $B_{z}$, even $m$'s occur in an oscillatory fashion. The larger even value of $m$ is, the weaker strength of $B_{z}$ is required. The oscillation begins around the energy of saddle point ($E^{c}\approx0.028$ eV; $E^{v}\approx-0.035$ eV), where the D-group LLs are turned into pairs of entangled LLs. The alternating appearance of odd and even $m$'s at low $E_{F}$ reveals the strong competitive and cooperative relations between the intralayer and interlayer atomic interactions and the Coulomb potentials.

The peculiar change of $m$ is also seen in the $E_{z}-E_{F}-\sigma_{xy}$ phase diagram but with distinct features, as shown in Fig. 6(b). The zero plateau ($m=0$) exists at $E_{z}<0.302$, where the other higher plateaus obey the usual integer QHE, i.e., $m=1, 2, 3,4$, etc. Increasing $E_{z}$ diminishes the width of zero plateau and raises $m$ for a chosen $E_{F}$. In addition, beyond a critical value of $E_{z}$ (indicated by the white arrow), the plateaus with an even $m$ gradually turn into the oscillatory modes and finally disappear for further stronger $E_{z}$. At the same time, the half-integer QHE in units of $4e^{2}/h$ is formed. The oscillatory behavior of $\sigma_{xy}$ gets stronger away from $E_{F}=0$, corresponding to the more frequent LL anticrossings/crossings. The two phase plots suggest that one may increase $B_{z}$/$E_{z}$ strength to lift/enhance the LL degeneracy, and to lead to the crossover from integer/half-integer to half-integer/integer QHEs. This provides an alternative to possible transport applications.

\bigskip
\bigskip
\centerline {\textbf {IV. Concluding Remarks}}%
\bigskip
\bigskip

The strong electrically tunable energy bands are capable of enriching the quantization properties under a uniform perpendicular magnetic field ${B_z\hat z}$, such as producing two subgroups of Landau levels (LLs), uniform and non-uniform LL energy spacings, and frequent crossings and anti-crossings. Consequently, the magnetotransport properties of bilayer phosphore cover the integer and half-integer conductivities, the well-like, staircase and composite quantum structures, as well as different step heights and widths. The rich composite effects of intrinsic interactions and external fields are responsible for this. $E_{F}-B_{z}-\sigma_{xy}$ and $E_{z}-E_{F}-\sigma_{xy}$ phase plots reveal the transitions between insulator to conductor, integer to half-integer QHEs, and the oscillating quantized numbers.

Our predictions for the diverse quantum conductivities could be experimentally verified by magneto-transport measurements in a dual-gated system, as have been carried out on few-layer graphenes \cite{Lee2013,Velasco2012,Weitz2011}. Phosphorene is anticipated to be an attractive candidate for designing unconventional thermoelectric devices \cite{Fei2014,LingX2015}. These results could be further used to understand the magnetothermoelectric transport. In addition, the methodology is readily extended to other main-stream 2D systems, such as few-layer silicene, germanene, tinene, antomonene, bismuthene, etc. We sincerely believe that our systematic study could attract more experimental/theoretical researches on these emergent materials and accelerate their developments for device and energy applications.

\centerline {\textbf {ACKNOWLEDGMENT}}%

\bigskip

\bigskip

\noindent
This work was supported by the MOST of Taiwan, under Grant No. MOST 107-2112-M-992-001.

\par\noindent ~~~~$^\star$e-mail address: yarst5@gmail.com

\bigskip \vskip0.6 truecm
\newpage

\begin{figure}
\centering
\includegraphics[width=0.85\textwidth]{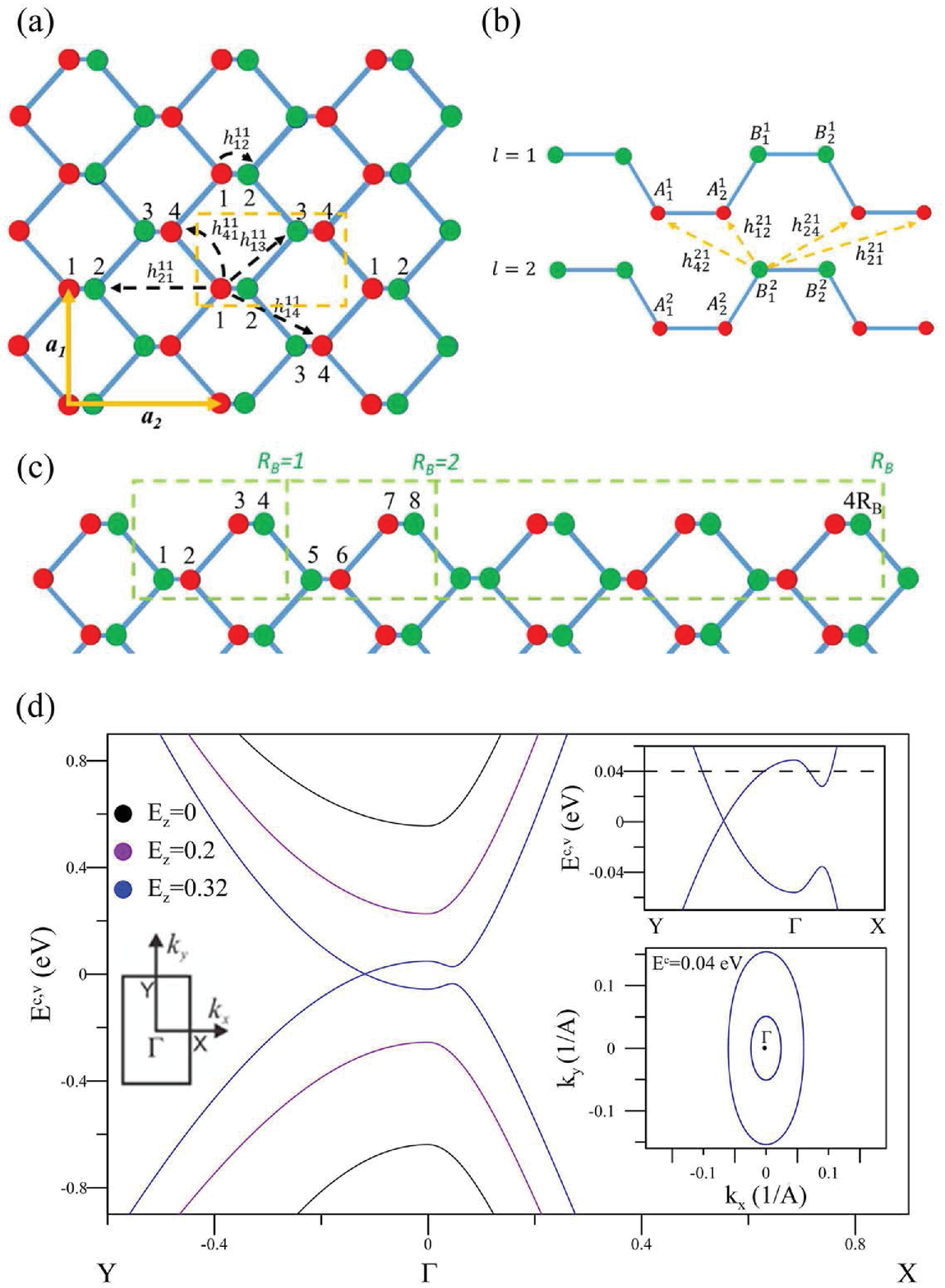}
\caption{The top view for the geometric structure of monolayer phosphorene in (a), and the side view for that of bilayer phosphorene in (b), with various intralayer and interlayer atomic interactions. In the presence of a uniform perpendicular magnetic field, an enlarged unit cell is in rectangular shape as shown in (c) by the dashed green lines. The band structures of bilayer phosphorene under various electric fields are presented in (d).
}
\label{figure:1}
\end{figure}

\begin{figure}
\centering
\includegraphics[width=0.7\textwidth]{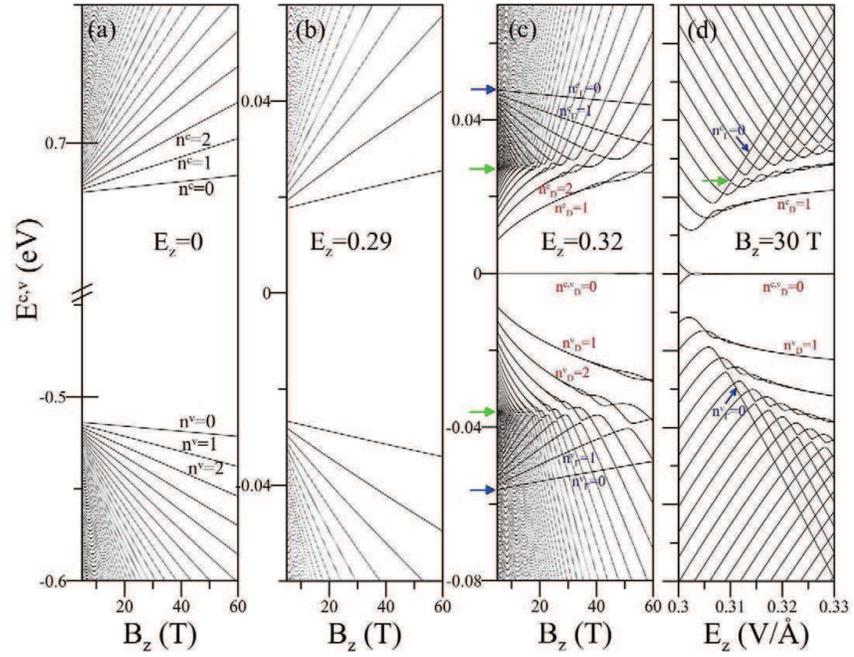}
\caption{(a)-(c) The $B_{z}$-dependent LL spectra for various $E_{z}$'s and (d) the $E_{z}$-dependent LL spectrum at $B_{z}$=30 T.}
\label{figure:2}
\end{figure}

\begin{figure}
\centering
\includegraphics[width=0.7\textwidth]{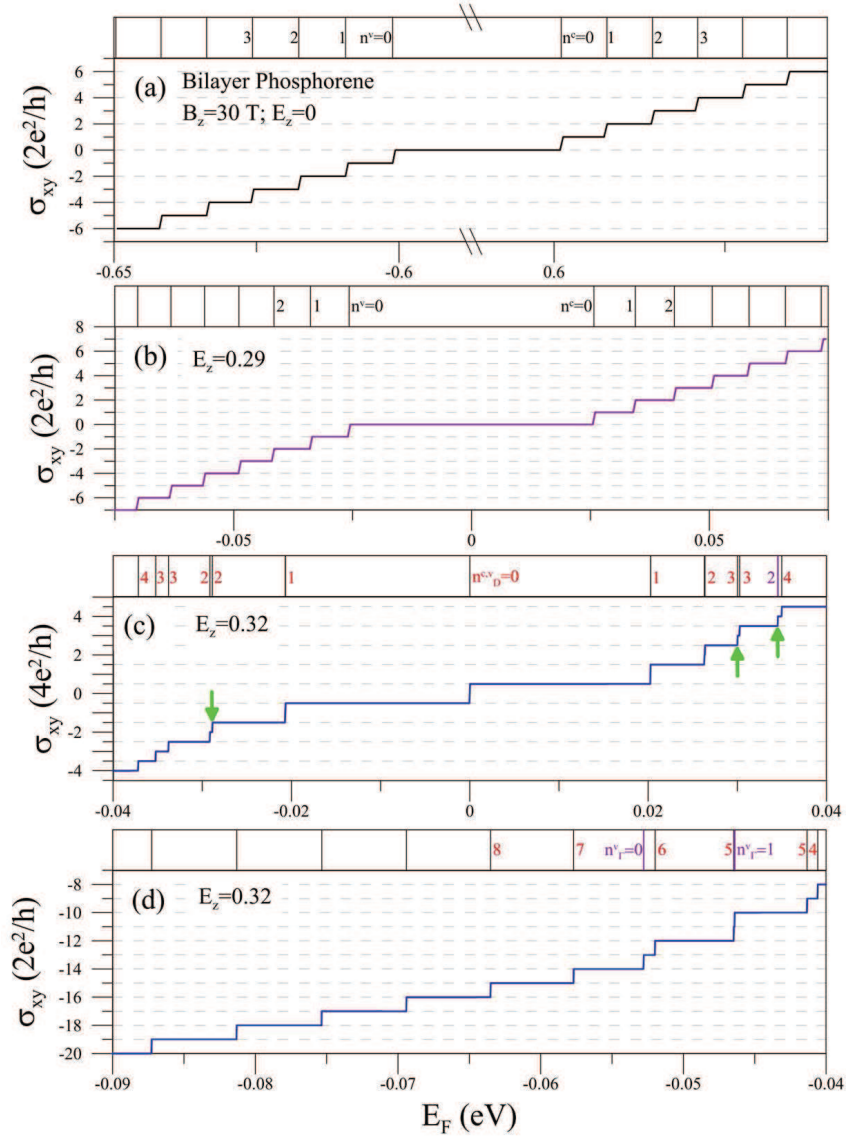}
\caption{The $E_{F}$-dependent transverse conductivity for various $E_{z}$'s at $B_{z}$=30 T. The upper inset in each panel shows the corresponding LL energies.}
\label{figure:3}
\end{figure}

\begin{figure}
\centering
\includegraphics[width=0.7\textwidth]{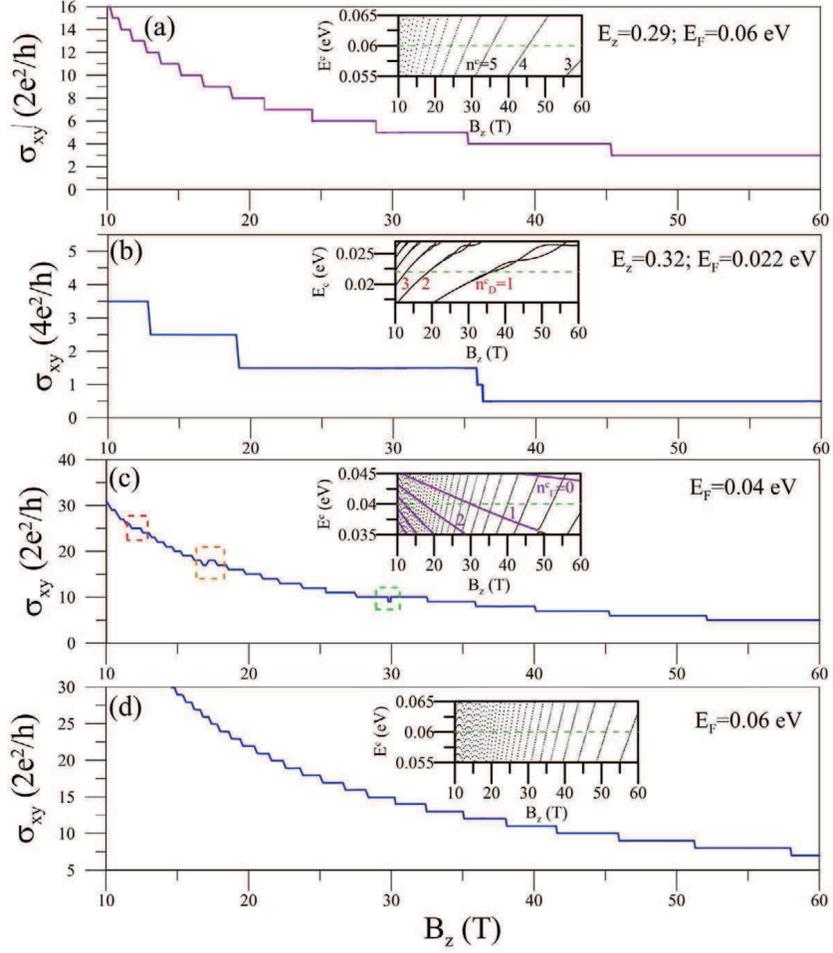}
\caption{(a) The $B_{z}$-dependent transverse conductivity at $E_{z}=0.29$ $V/{\AA}$ and $E_{F}=0.06$ eV. The same plots at $E_{z}=0.32$ $V/{\AA}$ for various $E_{F}$'s are shown in (b)-(d). Also shown in the insets are the $B_{z}$-dependent LL energy spectra, with the Fermi levels represented by the dashed green lines.}
\label{figure:4}
\end{figure}

\begin{figure}
\centering
\includegraphics[width=0.7\textwidth]{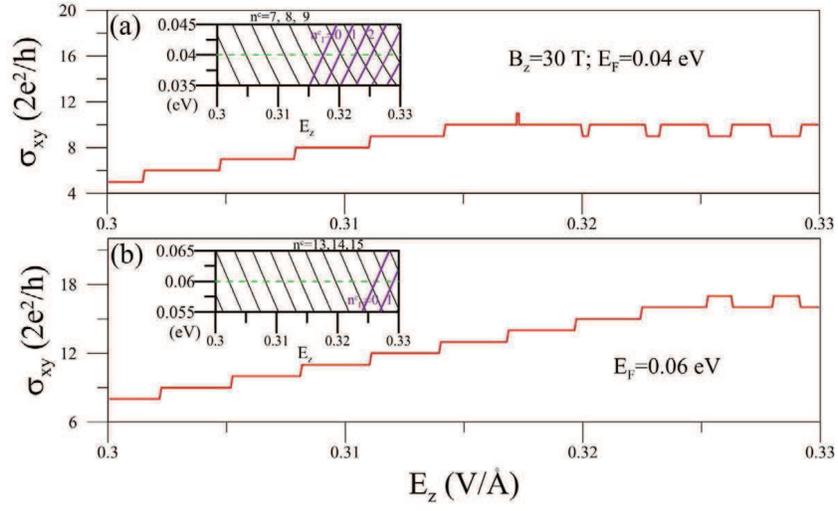}
\caption{The $E_{z}$-dependent $\sigma_{xy}$ at $B_{z}$=30 T for different $E_{F}$'s in (a) and (b). The corresponding $E_{z}$-dependent LL energy spectra are shown in the insets.}
\label{figure:5}
\end{figure}

\begin{figure}
\centering
\includegraphics[width=0.7\textwidth]{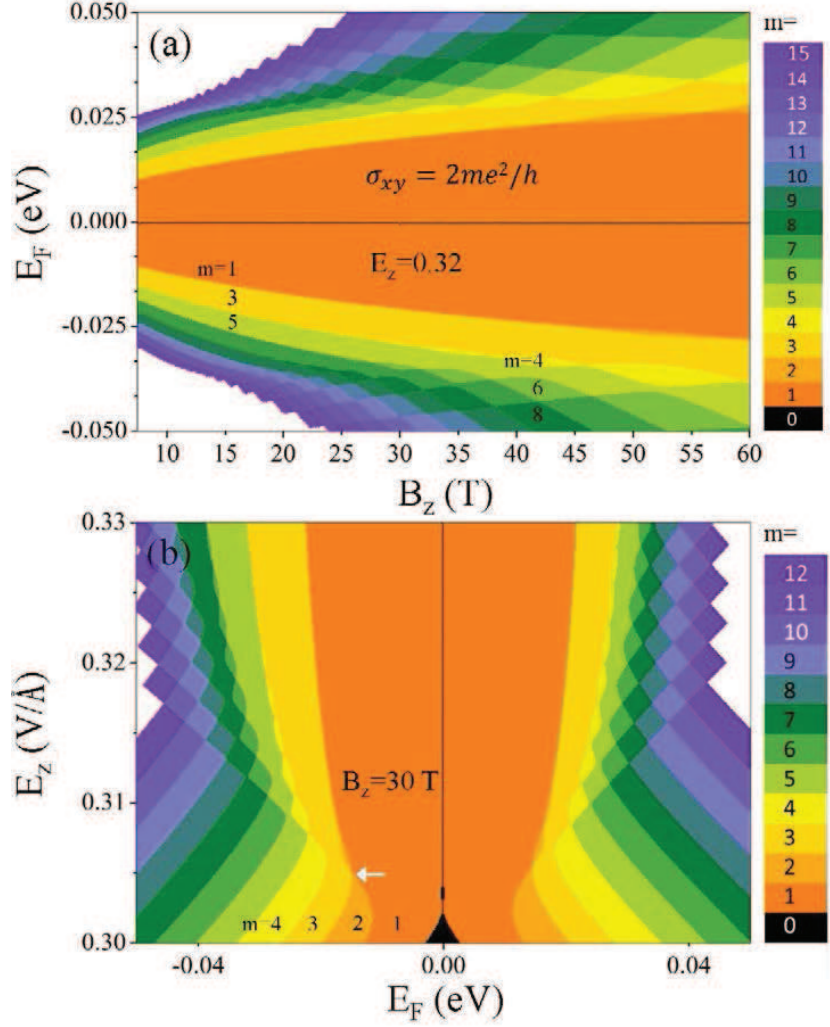}
\caption{The $E_{F}-B_{z}-\sigma_{xy}$ phase diagram in (a) and that of $E_{z}-E_{F}-\sigma_{xy}$ in (b). The color scale represents the value of quantized number $m$, where $m\leq 15$ in (a) and $m\leq 12$ in (b). Also, the white arrow indicates the minimum value of $E_{z}$ to induce the oscillation mode of even plateaus.}
\label{figure:6}
\end{figure}

\end{document}